\documentclass[a4paper,12pt]{article}

\usepackage[utf8]{inputenc}
\usepackage{cancel}
\usepackage{ulem}
\usepackage{amsfonts}
\usepackage{amssymb}
\usepackage{graphicx}
\usepackage{amsmath}
\usepackage{enumerate}
\usepackage{subfig}
\usepackage{color}
\usepackage{float}
\usepackage{cite}
\usepackage{mathrsfs}
\usepackage{epsfig}
\usepackage{dcolumn}
\usepackage{multirow}
\usepackage{placeins}
\usepackage{bm}
\usepackage{amsmath,amssymb,amsthm}
\usepackage[table]{xcolor}
\usepackage[colorlinks=true,linkcolor=blue]{hyperref}
\hypersetup{citecolor=magenta}

\title{\bf \Large{Thermodynamic stability and geometric thermodynamics of charged black holes in Rastall--massive gravity under quintessence}
}

\author{Mohamed Chabab$^{1,2}$\footnote{mchabab@uca.ac.ma  } ,
	\  Samir Iraoui$^{1,3}$\footnote{samir.iraoui@ced.uca.ma (Corresponding author)},
	\ Hicham Sriba$^1$\footnote{h.sriba.ced@uca.ac.ma}\\
	\\ 
	{\small $^1$ Cadi Ayyad University, Faculty of  Science Semlalia, LPHEAG, 
	}\\
	{\small  P.O.B. 2390 Marrakech 40000, Morocco.
	}\\
	{\small  $^2$ Cadi Ayyad University, National School of Applied Sciences,  P.O.B.  63, } \\
	{ \small  Safi 46000, Morocco} \\ 
	{\small  $^3$ Centre Regional des Metiers de l'Education et de la Formation de Marrakech,}\\
	{ \small  Marrakech 40000, Morocco} 
} 

\date{}

\begin{document} 
	\maketitle
	
	\begin{abstract}
In the present paper, we derive a new charged black hole solution embedded in a quintessence field within  a framework combining  the Rastall and massive gravity in an asymptotically flat spacetime.  We interpret the integration constant associated with the structural properties of the quintessence field as a thermodynamical variable. In this context,  several limiting cases have been explored with the aim to shed light on the role of the model parameters. Then, we analyse the thermodynamic properties and stability of the charged black holes in flat spacetime, where the quintessence field is characterized by the state parameter \( \omega_q = -2/3 \). Detailed studies are perfomed to see how the various parameters affect phase transitions between small and large black holes. In particular, we establish a connection between the impact parameter and unstable circular photon orbits with the thermodynamic phase transitions. Our analysis  suggest that the photon sphere observables could remarkably  provide an indirect observational signature of black hole thermodynamic critical transitions.  Generally, thanks to the effective contribution of the quintessence parameter, we found that the overall thermodynamic behavior basically resembles that of the charged RN--AdS black hole with thermal features analogous to of a Van der Waals fluid ones. Finally, we have checked these findings using  thermodynamic geometry with the Quevedo metric.  Our results showed that the critical behaviour derived from the standard thermodynamics is fully consistent with geothermodynamics description, where the phase structure and divergence points exactly coincide with those identified from heat capacity and Gibbs free energy analyses.
	\end{abstract}

	{\small\itshape Keywords: Charged black hole; thermodynamics; geothermodynamics; massive gravity; Rastall gravity; unstable photon orbit; quintessence.}
	
	\tableofcontents
	
	\renewcommand{\thefootnote}{\arabic{footnote}}
	\setcounter{footnote}{0}
	
	\newpage
	\section{Introduction}
	\label{intro}
	
	Black holes are among the most fascinating predictions of General Relativity (GR), providing a unique framework to test gravitational physics in the strong-field regime. Over the years, a wide variety of black hole solutions have been constructed, revealing deep connections between geometry and physics \cite{Hawking:1975vcx,Bardeen:1973gs}. Despite its success, GR faces several challenges, particularly in explaining cosmological observations such as the accelerated expansion of the Universe. These issues have motivated the development of modified theories of gravity. Among them, Rastall gravity proposes a deviation from the standard conservation law of the energy-momentum tensor, allowing its divergence to be proportional to the gradient of the Ricci scalar \cite{Rastall:1972swe,Rastall:1976uh}. This modification introduces a non-minimal coupling between matter and geometry and leads to novel black hole solutions with distinct physical properties \cite{Lobo:2017dib,Heydarzade:2017wxu}.
	
	In parallel, massive gravity has emerged as another important extension of GR, in which the graviton acquires a finite mass \cite{Fierz:1939ix,deRham:2010ik, deRham:2010kj}. This framework modifies the large-scale behavior of gravity and provides a possible explanation for cosmic acceleration without invoking dark energy explicitly \cite{Akrami:2012vf,Akrami:2015qga}. In particular, the ghost-free formulation of massive gravity allows for consistent black hole solutions with rich thermodynamic behavior \cite{Cai:2014znn,Hendi:2018xuy,Hendi:2015hoa,Hendi:2015pda,Ghosh:2015cva}. The interplay between massive gravity and other modified theories, such as Rastall gravity, opens new avenues for exploring gravitational physics beyond the standard framework.
	
	Black hole thermodynamics establishes a profound connection between gravitation, quantum theory and statistical mechanics. Discovering that black holes possess temperature and entropy has led to the formulation of thermodynamic laws analogous to those of ordinary systems \cite{Bekenstein:1973ur,Jacobson:1995ab,Kastor:2009wy, Dolan:2011xt}. In the extended phase space approach, the cosmological constant is interpreted as a thermodynamic pressure, giving rise to a rich phase structure. In particular, charged AdS black holes exhibit phase transitions analogous to those of a Van der Waals fluid, characterized by critical behavior and universal ratios \cite{Dolan:2012jh,Rajagopal:2014ewa,Kubiznak:2012wp,Kubiznak:2016qmn}. The inclusion of additional fields, such as quintessence,  enriches further this structure by modifying both the geometry and thermodynamic properties of the black holes \cite{Kiselev:2002dx,Li:2014ixn}.
	
	An alternative perspective on thermodynamics is provided by geothermodynamics, where thermodynamic systems are described using differential geometry \cite{Hendi:2015rja,Weinhold:1975xej,Ruppeiner:1995zz,Bhattacharya:2017hfj}. In this approach, the curvature scalar associated with a thermodynamic metric encodes information about interactions and phase transitions. The Quevedo formalism, based on Legendre invariance, provides a consistent geometric description in which curvature singularities generally coincide with thermodynamic critical points \cite{Quevedo:2006xk,Quevedo:2008xn}. This method has been successfully applied to a wide range of black hole solutions, offering a complementary way to analyze thermodynamical behavior \cite{Hendi:2015xya,Chabab:2019mlu,Soroushfar:2019ihn}.
	
	More recently, a remarkable connection has been established between black hole thermodynamics and the properties of null geodesics, particularly unstable circular photon orbits. It has been shown that the photon sphere radius and the associated critical impact parameter can encode information about thermodynamic phase transitions \cite{Chandrasekhar:2018sjg}. These quantities exhibit behavior analogous to thermodynamic variables and are capable of reproducing the same critical structure observed in traditional thermodynamic analyses \cite{Wei:2017mwc,Chabab:2019kfs,NaveenaKumara:2019nnt}. This correspondence provides a new geometrical and potentially observational probe of black hole phase transitions.
	
	Motivated by these developments, it is natural to investigate black hole solutions within the combined framework of Rastall and massive gravity in the presence of a quintessence field. Such a setup allows one to explore the interplay between modified gravity effects, surrounding matter fields, and thermodynamic properties. For related studies, see Ref.~\cite{Chabab:2020ejk,Thomas:2012zzc}. The presence of quintessence, characterized by a negative equation of state parameter, plays a crucial role in shaping both the spacetime geometry and the associated thermodynamic behavior. These modifications can significantly affect both the causal structure and the thermodynamic properties of black hole solutions.
	
	In this work, we derive a charged black hole solution in Rastall--massive gravity surrounded by a quintessence field and analyze its thermodynamic properties. We interpret the integration constant associated with the quintessence field as a thermodynamic variable and investigate the corresponding phase structure. Furthermore, we establish a connection between thermodynamic phase transitions and unstable circular photon orbits by analyzing the photon sphere radius and the critical impact parameter. Finally, we explore the thermodynamic geometry using the Quevedo metric, confirming the consistency between geometric and thermodynamic descriptions. The novelty of this work lies in providing the first unified framework combining Rastall gravity, massive gravity, with quintessence effects, their in establishing a consistent correspondence between thermodynamic criticality, photon sphere observables, and thermodynamic geometry.

\section{Charged black hole solutions surrounded by a quintessence field in Rastall--massive gravity}\label{sec2}

Rastall gravity modifies general relativity by relaxing the covariant conservation condition \( \nabla_\nu T^{\mu\nu} = 0 \). Instead, it introduces a generalized conservation law, which is expressed as follows \cite{Rastall:1972swe,Rastall:1976uh}:

\begin{equation}
\nabla_\nu T^{\mu\nu} = a^\mu. 
\end{equation}
To preserve consistency with the geometric structure of the theory, \( a^\mu \) is defined in terms of the Ricci scalar \( R \) as
\begin{equation}
a^\mu = \lambda \nabla^\mu R,
\end{equation}
where \( \lambda \) is the Rastall parameter. 

In the presence of massive gravity, the Einstein field equations can be written as
\begin{equation}
G_{\mu\nu} + \kappa \lambda g_{\mu\nu} R + m^2 \chi_{\mu\nu} = \kappa T_{\mu\nu},
\label{fieldequ}
\end{equation}
where \( \kappa \) denotes the gravitational coupling constant in Rastall theory, and \( m \) represents the graviton mass. The tensors \( G_{\mu\nu} \), \( T_{\mu\nu} \), and \( \chi_{\mu\nu} \) correspond to the Einstein tensor, the energy-momentum tensor, and the massive gravity contribution, respectively. The latter is defined as follows:

\[
\chi_{\mu\nu}= -\frac{c_1}{2}\, (\mathcal{U}_1 \,g_{\mu\nu} - \mathcal{K}_{\mu\nu}) - \frac{c_2}{2}\,(\mathcal{U}_2 \,g_{\mu\nu} - 2\,\mathcal{U}_1 \,\mathcal{K}_{\mu\nu} + 2\, \mathcal{K}^2_{\mu\nu})
\]
\[
-\frac{c_3}{2} \,(\mathcal{U}_3 \,g_{\mu\nu} - 3\,\mathcal{U}_2 \,\mathcal{K}_{\mu\nu} + 6\,\mathcal{U}_1\, \mathcal{K}^2_{\mu\nu} - 6\, \mathcal{K}^3_{\mu\nu})
\]
\begin{equation}
-\frac{c_4}{2} (\mathcal{U}_4\, g_{\mu\nu} - 4\,\mathcal{U}_3 \,\mathcal{K}_{\mu\nu} + 12\,\mathcal{U}_2 \,\mathcal{K}^2_{\mu\nu} - 24\,\mathcal{U}_1 \,\mathcal{K}^3_{\mu\nu} + 24 \,\mathcal{K}^4_{\mu\nu})
\label{tensormassive}
\end{equation}
The quantities \( \mathcal{U}_i \) denote the symmetric polynomials of the eigenvalues of the \( 4 \times 4 \) matrix defined by
\begin{equation}
\mathcal{K}_{\mu\nu} = \sqrt{g^{\mu\alpha} f_{\alpha\nu}}.
\end{equation}

Following \cite{Cai:2014znn,Chabab:2025jfi,Hendi:2015hoa}, we adopt the reference metric in four dimensions as
\begin{equation}
f_{\mu\nu} = \text{diag}(0, 0, c_0^2 h_{ij}),
\label{metricref}
\end{equation}
where \( c_0 \) is a positive constant. According to Eq.~\eqref{metricref}, the components of \( \mathcal{U}_i \) take the form \cite{Hendi:2015pda,Xu:2015rfa}
\begin{equation}
\mathcal{U}_1=\frac{2c_0}{r}, \qquad \mathcal{U}_2=\frac{2c_0^2}{r^2}, \qquad \mathcal{U}_3=\mathcal{U}_4=0.
\end{equation}
According to \cite{Yue:2024rwj}, the tensor \( \chi_{\mu\nu} \) given in Eq.~\eqref{tensormassive} can be written explicitly as
\begin{equation}
\chi^1_{\ 1}=\chi^2_{\ 2} = \frac{c_0 c_1}{r} + \frac{c_0^2 c_2}{r^2}, \qquad 
\chi^3_{\ 3}=\chi^4_{\ 4} = \frac{c_0 c_1}{2r}.
\end{equation}
	
We now consider a static, spherically symmetric spacetime in four dimensions \cite{Lobo:2017dib}, described by the metric

\begin{equation}
ds^2 = -f(r)\, dt^2 + \frac{dr^2}{f(r)} + r^2 \, d\Omega^2,
\end{equation}      

where \( f(r) \) is the metric function depending on the radial coordinate \( r \), and \( d\Omega^2 = d\theta^2 + \sin^2\theta \, d\phi^2 \) denotes the line element on the unit 2-sphere.

Using the field equations given in Eq.~\eqref{fieldequ}, we define the Rastall--massive tensor as
\begin{equation}
\Theta_{\mu\nu} = G_{\mu\nu} + \kappa \lambda g_{\mu\nu} R + m^2 \chi_{\mu\nu}.
\end{equation}
The non-vanishing components of the field equations can then be written as
\[
\Theta^{0}_{\ 0} = \frac{1}{r^2} \left[ r f'(r) + f(r) - 1 \right] + \beta R 
+ m^2 \left( \frac{c_0 c_1}{r} + \frac{c_0^2 c_2}{r^2} \right),
\]

\[
\Theta^{1}_{\ 1} = \frac{1}{r^2} \left[ r f'(r) + f(r) - 1 \right] + \beta R 
+ m^2 \left( \frac{c_0 c_1}{r} + \frac{c_0^2 c_2}{r^2} \right),
\]

\[
\Theta^{2}_{\ 2} = \frac{1}{r^2} \left[ r f'(r) + \frac{1}{2} r^2 f''(r) \right] 
+ \beta R + m^2 \frac{c_0 c_1}{2r},
\]

\begin{equation}
\Theta^{3}_{\ 3} = \frac{1}{r^2} \left[ r f'(r) + \frac{1}{2} r^2 f''(r) \right] 
+ \beta R + m^2 \frac{c_0 c_1}{2r}.
\end{equation}
Here, we have introduced the parameter \( \beta = \kappa \lambda \), which characterizes the deviation from standard energy-momentum conservation in Rastall gravity.

The Ricci scalar is given by
\begin{equation}
R = -\frac{1}{r^2} \left[ r^2 f''(r) + 4r f'(r) + 2 f(r) - 2 \right].
\label{Ricci}
\end{equation}
In this framework, the total energy-momentum tensor \( T_{\mu\nu} \) is written as
\begin{equation}
T_{\mu\nu} = E_{\mu\nu} + \mathcal{T}_{\mu\nu},
\label{tenergy}
\end{equation}
where the first term corresponds to the electromagnetic contribution. The non-vanishing components of the Maxwell tensor \( E^\mu_{\ \nu} \) are given by
\begin{equation}
E^\mu_{\ \nu}=\frac{Q^2}{\kappa\,r^4}\, \mathrm{diag}(-1,\,-1,\,1,\,1).
\end{equation}
Moreover, the second term appearing in Eq.~\eqref{tenergy} represents the energy-momentum tensor of the surrounding field, whose non-vanishing components are given by \cite{Kiselev:2002dx,Heydarzade:2017wxu}
\begin{equation}
\mathcal{T}^0_{\ 0}=\mathcal{T}^1_{\ 1}=-\rho_s, \qquad 
\mathcal{T}^2_{\ 2}=\mathcal{T}^3_{\ 3}=\frac{1}{2}(1+\omega_s)\rho_s.
\end{equation}
Here, \( \rho_s \) and \( \omega_s \) denote the energy density and the equation-of-state parameter, respectively.

Finally, by applying Eq.~\eqref{fieldequ} and imposing the equality of the non-vanishing components \( \Theta^0_{\ 0}=T^0_{\ 0} \) and \( \Theta^1_{\ 1}=T^1_{\ 1} \), we obtain the following differential equation:

\begin{multline}
\frac{1}{r^2} \left[ r f'(r) + f(r) - 1 \right]  
- \frac{\beta}{r^2} \left[ r^2 f''(r) + 4r f'(r) + 2 f(r) - 2 \right]\\
+ m^2 \left(\frac{c_0 c_1}{r} + \frac{c_0^2 c_2}{r^2}\right)
= -\left(\frac{Q^2}{r^4}+\kappa\rho_s\right).
\label{difeq1}
\end{multline}
Similarly, from the angular components \( \Theta^2_{\ 2}=T^2_{\ 2} \) and \( \Theta^3_{\ 3}=T^3_{\ 3} \), we obtain
\begin{multline}
\frac{1}{r^2} \left[ r f'(r) + \frac{1}{2} r^2 f''(r) \right] 
- \frac{\beta}{r^2} \left[ r^2 f''(r) + 4r f'(r) + 2 f(r) - 2 \right]\\
+ m^2 \frac{c_0 c_1}{2r}
= \frac{1}{2}(1+\omega_s)\kappa\rho_s + \frac{Q^2}{r^4}.
\label{difeq2}
\end{multline}
We thus obtain a system of two differential equations involving the unknown functions \( f(r) \) and \( \rho_s \). Eliminating \( \rho_s \) between Eqs.~\eqref{difeq1} and \eqref{difeq2} allows one to determine the metric function \( f(r) \), after which \( \rho_s \) can be obtained by substitution.

For a quintessence field, we consider the state parameter \( \omega_q = -\frac{2}{3} \) \cite{Li:2014ixn,Kiselev:2002dx}. a standard choice in the literature that lies strictly within the physical quintessence regime $-1 < \omega_q < -1/3$. This choice renders the metric function linear in $r$ for the dark energy component, allowing for exact analytical determinations of the critical parameters.

The general solution for the metric function takes the form
\begin{equation}
f(r)=1-\frac{2M}{r}+\frac{Q^2}{r^2} + \alpha\,r^{\frac{1+2\beta}{1-\beta}}+ a\,r+ b,
\label{metric}
\end{equation}
where \( M \) and \( Q \) denote the mass and electric charge of the black hole, respectively. The parameter \( \alpha \) is an integration constant associated with the surrounding quintessence field, while the quantities \( a \) and \( b \) encode the contributions of massive gravity, defined as
\begin{equation}
a = m^2 \frac{c_0 c_1}{2}, \qquad b = m^2 c_0^2 c_2.
\end{equation}
The additional terms arising from Rastall and massive gravity contributions modify the effective gravitational potential and the asymptotic behavior of the spacetime. 

In the limiting case \( \beta = \frac{1}{4} \), the solution reduces to a  charged RN--AdS black hole in massive gravity, where \( \alpha \propto 1/l^2 \), with \( l \) being the AdS radius, corresponding to the cosmological constant \( \Lambda = -3/l^2 \),
\begin{equation}
f(r)=1-\frac{2M}{r} + \frac{Q^2}{r^2} + \alpha r^2 + a r + b.
\end{equation}
In the absence of both the quintessence field and massive gravity contributions (i.e., \( \alpha = 0 \) and \( a=b=0 \)), the solution is actually the standard Reissner--Nordström black hole,
\begin{equation}
f(r)=1-\frac{2M}{r} + \frac{Q^2}{r^2}.
\end{equation}

For \( \beta = 0 \), the metric simplifies to
\begin{equation}
f(r)=1-\frac{2M}{r} + \frac{Q^2}{r^2} + (\alpha + a)\,r + b,
\end{equation}
which describes a charged black hole in massive gravity surrounded by a matter field with a linear radial dependence.
For \( \beta = -\frac{1}{2} \), one readily obtains,
\begin{equation}
f(r)=1-\frac{2M}{r} + \frac{Q^2 + \alpha}{r^2} + a\,r + b,
\end{equation}
which can be interpreted as a charged solution with an effective charge \( Q_{\mathrm{eff}}^2 = Q^2 + \alpha \).
For \( \beta = -2 \), the metric becomes
\begin{equation}
f(r)=1-\frac{2M + \alpha}{r} + \frac{Q^2 }{r^2} + a\,r + b,
\end{equation}
corresponding to an effective mass \( M_{\mathrm{eff}} = M + \frac{\alpha}{2} \).

From Eq.~\eqref{Ricci}, the Ricci scalar associated with the metric function can be evaluated, one straightforwardly finds that its leading behavior is given by
\begin{equation}
R \sim \frac{1}{r} + \frac{\alpha (\beta +2)}{(\beta - 1)^2} r^{-\frac{5\beta - 2}{\beta -1}}.
\label{ExpRicci}
\end{equation}
The first term diverges as \( r \to 0 \), while the second term may also diverge depending on the value of \( \beta \), namely for \( \beta \leq \frac{1}{4} \) or \( \beta > 1 \). Therefore, the Ricci scalar diverges at \( r=0 \), indicating a genuine curvature singularity, while it remains finite for \( r>0 \).
	
\begin{figure}[h]
	\centering  
	\includegraphics[width=0.32\linewidth]{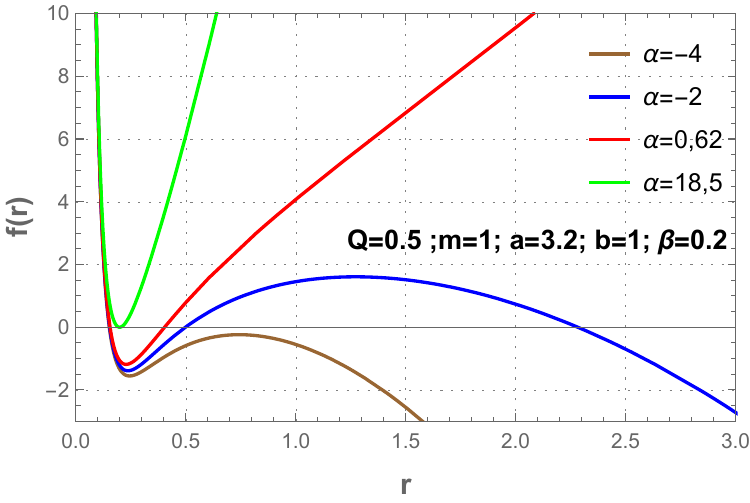}  
	\includegraphics[width=0.32\linewidth]{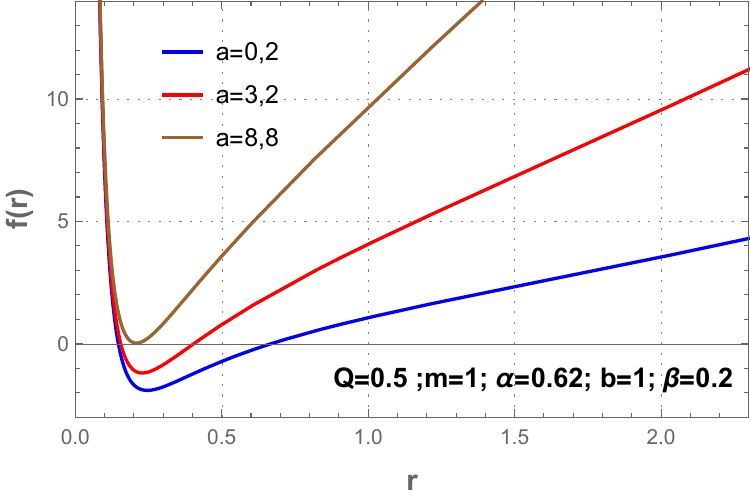} 
	\includegraphics[width=0.32\linewidth]{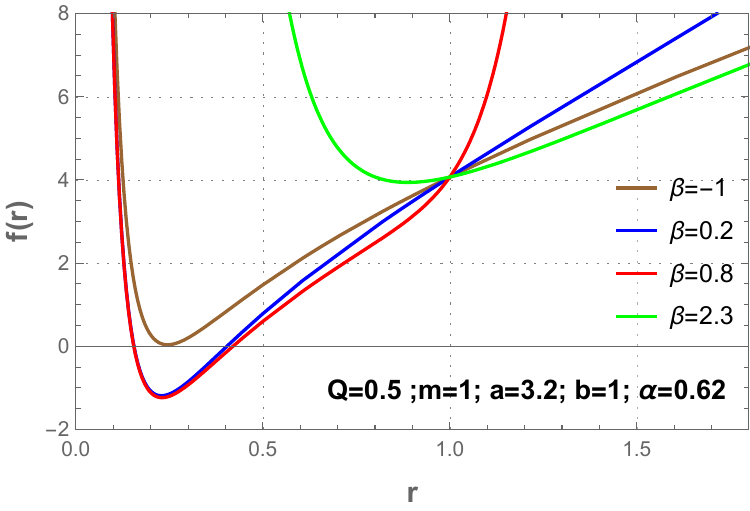}   
	\caption{	Behavior of the metric function \( f(r) \) as a function of the radial coordinate \( r \) for different values of the parameters \( \alpha \), \( a \), and \( \beta \), illustrating the modification of the horizon structure.}
	\label{fmetric}
\end{figure}

In Fig.~\ref{fmetric}, we illustrate the behavior of the metric function \( f(r) \), given in Eq.~\eqref{metric}, as a function of the radial coordinate \( r \) for different values of the model parameters, in order to analyze their impact on the spacetime structure.
As shown in Fig.~\ref{fmetric}, the zeros of the metric function \( f(r) \) correspond to the horizons of the black hole spacetime. Depending on the values of the parameters \( \beta \), \( \alpha \), and \( a \), the solution may exhibit one or two horizons, which can be interpreted as the event horizon and, when present, a cosmological horizon.
The variation of these parameters significantly affects both the number and the location of the horizons, reflecting the influence of the Rastall parameter, the quintessence field, and the massive gravity contributions on the global structure of the spacetime.

For large values of \( r \), the metric function behaves as \( f(r) \to \pm \infty \), depending on the dominant terms. In the limit \( r \to 0 \), one finds \( f(r) \to +\infty \), indicating the presence of a curvature singularity at the origin.

Having obtained the solution for a charged black hole surrounded by a quintessence field within the framework of Rastall--massive gravity, we now proceed to compute the thermodynamic quantities in order to analyze the effects of the quintessence field on phase transitions.

The mass is determined from the horizon condition \( f(r_h)=0 \), and is given by
\begin{equation}
M(r_h,\alpha)=\frac{ Q^2 + r_h^2 (1 + b + a r_h) + \alpha\, r_h^{\frac{3}{\beta -1}}}{2\, r_h}.
\label{mass}
\end{equation}
The Hawking temperature, related to the surface gravity, reads
\begin{equation}
T_H(r_h,\alpha)=\frac{1}{4\pi}\left(2a + \frac{1 + b}{r_h} - \frac{Q^2}{r_h^3} - \alpha \frac{\beta +2}{\beta - 1} r_h^{-\frac{3\beta}{\beta -1}}\right).
\label{temp}
\end{equation}
From Eq.~\eqref{temp}, the integration parameter \( \alpha \), associated with the structural properties of the black hole's quintessence field, can be expressed in terms of the temperature \( T \), the horizon radius \( r_h \), and the Rastall parameter \( \beta \) as follows:
\begin{equation}
\alpha(r_h,T) = \frac{\beta -1}{\beta +2}\, r_h^{\frac{3}{\beta -1}}\,
\left[(1+b)\,r_h^2 - Q^2 + 2(a-2\pi T)\,r_h\right].
\label{eqalpha}
\end{equation}

 In the context of Rastall-massive gravity in presence of a quintessence field, the first law of extended thermodynamics \cite{Gibbons:1977mu} is expressed as:  
    \begin{equation}
        dM=TdS + \phi dQ + \Psi d\alpha + \sigma d\beta + \mathcal{A} da + \mathcal{B} db
        \label{firstlaw}
    \end{equation}
where T, \(\phi\), \(\Psi\), \(\sigma\), \(\mathcal{A}\) and \(\mathcal{B}\) are the conjugate quantities of the parameters S, Q, \(\alpha\), \(\beta\), a and b respectively. While the Smarr relation by invoking Euler's theorem \cite{Smarr:1972kt} is readily given by,
    \begin{equation}
        M = 2TS + \phi Q + \Psi \alpha + \sigma \beta + \mathcal{A} a + \mathcal{B} b
    \end{equation}
	
Next, we examine the thermodynamic criticality of the charged RN black hole in the framework of Rastall--massive gravity. To this end, we impose the critical conditions,
\[
\left.\frac{\partial T}{\partial r_h}\right|_{\alpha=\alpha_c,r_h=r_c}=0, 
\qquad 
\left.\frac{\partial^2 T}{\partial r_h^2}\right|_{\alpha=\alpha_c,r_h=r_c}=0.
\]
The corresponding critical quantities are given by
\begin{align}
r_c & = \frac{3 Q}{\sqrt{(1 + b) (1 + 2 \beta)}},\\
\alpha_c&=-\frac{2 Q^2 (\beta -1)^3 3^{\frac{3}{\beta-1}} 
	\left(\frac{3Q}{\sqrt{(1 + b) (1 + 2 \beta)}}\right)^{\frac{3}{\beta -1}}}
{\beta (2 + \beta) (1 + 2 \beta)},\label{criticalp}\\
T_c &=\frac{a}{2\pi} + \frac{\left[(1+b)(1+2\beta)\right]^{3/2}}{54\pi Q \beta}.
\end{align}

\begin{figure}[h]
	\centering  
	\includegraphics[scale=0.52]{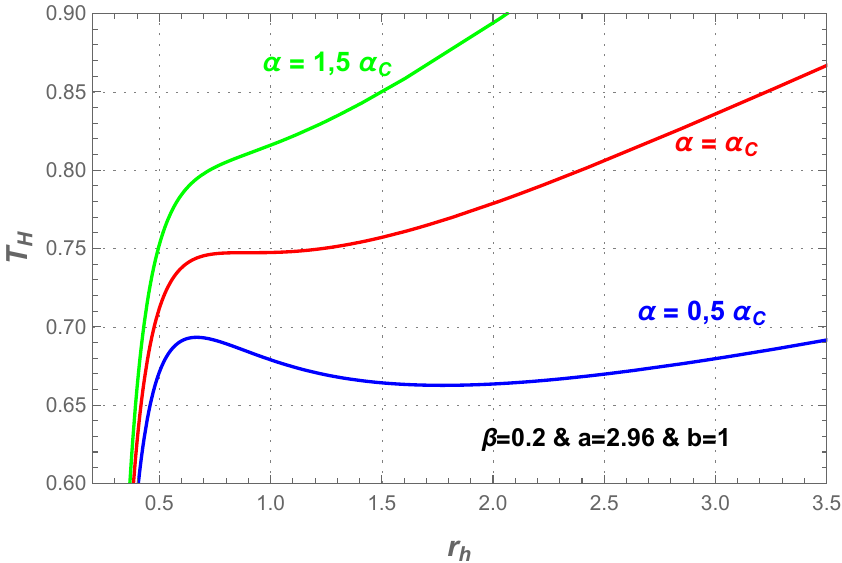}  
	\includegraphics[scale=0.52]{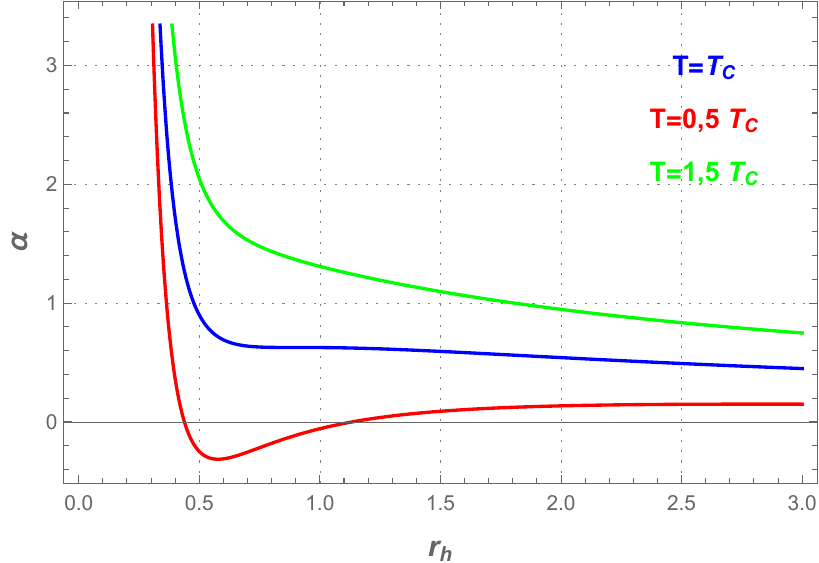}  
	\caption{Illustrating the small/large black hole phase transition. \textbf{Left:} Hawking temperature as a function of \( r_h \). \textbf{Right:} Quintessence parameter \( \alpha \) as a function of the horizon radius \( r_h \). We set \( b=1 \) and \( Q=0.5 \).}
	\label{alpha}
\end{figure}

In Fig.~\ref{alpha}, the right panel shows the behavior of the parameter \( \alpha \), associated with the quintessence field, as a function of the horizon radius. The left panel displays the Hawking temperature as a function of \( r_h \). In the figure, the curves exhibit a behavior analogous to that of a Van der Waals fluid \cite{Kubiznak:2012wp}, indicating that Eq.~\eqref{eqalpha} plays the role of an effective equation of state. This suggests that the parameter \( \alpha \) can be effectively interpreted as a thermodynamic parameter playing a role analogous to pressure, for fixed values of the Rastall parameter. The diagram \( T\text{--}r_h \)  exhibits an inflection point at \( \alpha=\alpha_c \), separating monotonic and non-monotonic behaviors corresponding to \( \alpha>\alpha_c \) and \( \alpha<\alpha_c \), respectively. This behavior signals a phase transition between small and large black hole phases.

Within the same framework, we now compute the Gibbs free energy \( G \) and the heat capacity \( C_Q \) in order to investigate the phase structure and thermodynamic stability,
\begin{align}
G(r_h,\alpha)&=\frac{1}{4}\left[r_h(1+b) + \frac{3Q^2}{r_h} + \frac{3\alpha\beta}{\beta -1}r_h^{\frac{2+\beta}{1-\beta}}\right],
\label{gibbs}\\
C_Q(r_h,\alpha)&=2\pi r_h^2 
\frac{(1+b)r_h^2 - Q^2 + 2a r_h^3 - \alpha \frac{2+\beta}{\beta - 1} r_h^{\frac{-3}{\beta -1}}}
{3Q^2 - (1+b)r_h^2}.
\label{Cp}
\end{align}
\begin{figure}[h]
	\centering  
	\includegraphics[scale=0.52]{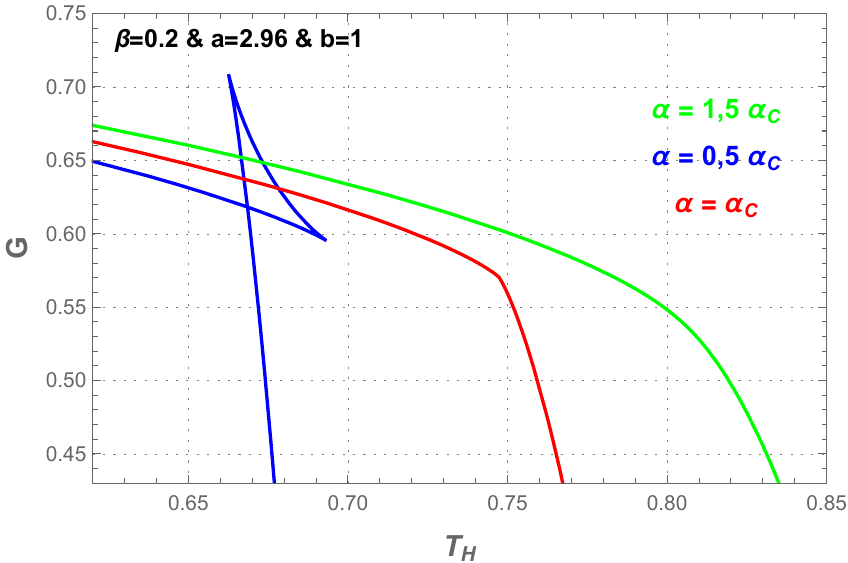}  
	\includegraphics[scale=0.52]{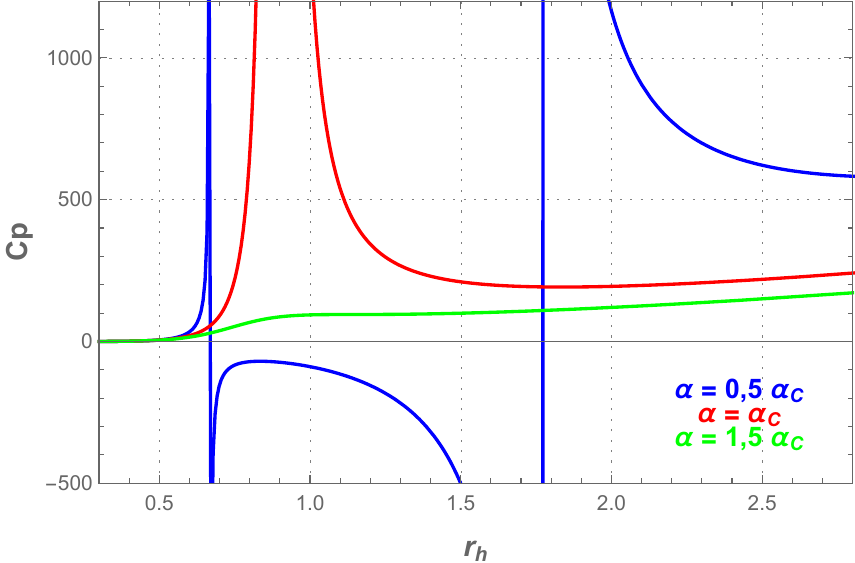}  
	\caption{\textbf{Left:} Gibbs free energy as a function of temperature where the swallowtail behavior signals a first-order phase transition. \textbf{Right:} Heat capacity as a function of the horizon radius. We set \( b=1 \) and \( Q=0.5 \).}
	\label{GCp}
\end{figure}

In the left panel of Fig.~\ref{GCp}, the Gibbs free energy exhibits a characteristic swallowtail behavior, indicating a first-order phase transition for \( \alpha < \alpha_c \). This behavior is analogous to that observed for the charged RN--AdS black hole in massive gravity at pressures below the critical value, which supports the interpretation of the parameter \( \alpha \) as an effective thermodynamic parameter playing a role analogous to pressure.
In the right panel of Fig.~\ref{GCp}, the thermodynamic stability shows the same qualitative behavior as that of the charged RN--AdS black hole in massive gravity \cite{Nam:2018ltb}, where the sign of the heat capacity distinguishes between stable and unstable phases.
In the following table, we present representative values of the Rastall parameter \( \beta \), together with the corresponding massive gravity parameter \( a \), in order to explore the validity domain of the model. The parameter \( b \) is fixed to 1, and the electric charge is set to \( Q=0.5 \). Note that the Rastall parameter is not defined for the values \( \beta = -2, -\frac{1}{2}, 0, 1 \).
\begin{center}
	\begin{tabular}{c c c c c c c}
		\hline \hline
		$ \beta $ & $ a $ & $r_c$ & $T_c$ & $\alpha_c$ & $\frac{\alpha_c \nu_c}{T_c}$ & Phase transition \\    
		\hline\hline
		0.06 & 20.87 & 1.00 & 3.98 & 2.98 & $\frac{3}{8}$ & present \\
		\hline
		0.10 & 10.31 & 0.96 & 2.08 & 1.61 & $\frac{3}{8}$ & present \\
		\hline
		0.15 & 5.29 & 0.93 & 1.17 & 0.94 & $\frac{3}{8}$ & present \\
		\hline
		0.20 & 2.69 & 0.89 & 0.85 & 0.63 & $\frac{3}{8}$ & present \\
		\hline
		0.25 & 1.68 & 0.86 & 0.51 & 0.44 & $\frac{3}{8}$ & present \\
		\hline
		0.30 & 0.90 & 0.83 & 0.37 & 0.33 & $\frac{3}{8}$ & present \\
		\hline
		0.35 & 0.40 & 0.81 & 0.27 & 0.25 & $\frac{3}{8}$ & present \\
		\hline
		0.40 & 0.075 & 0.79 & 0.21 & 0.20 & $\frac{3}{8}$ & present \\
		\hline
		$-5 < \beta < 0$ & $< 0$ & \dots & \dots & \dots & \dots & absent \\
		\hline
		$\beta > \frac{2}{5}$ & $< 0$ & \dots & \dots & \dots & \dots & absent \\
		\hline\hline
	\end{tabular}
	\captionof{table}{Analysis of the phase transition behavior in terms of the characteristic parameters of the model.}
	\label{table1}
\end{center}

For \( 0 < \beta \leq \frac{2}{5} \), the system exhibits first-order phase transitions analogous to those of a Van der Waals fluid. These additional contributions modify both the asymptotic structure of spacetime and the effective gravitational potential, leading to non-trivial thermodynamic behavior.
	
\section{Thermodynamic phase transition from unstable circular photon orbits}
\subsection{Geodesic equations of motion}

In this section, we investigate the motion of photons in the background spacetime described by the metric given in Eq.~\eqref{metric}. The dynamics are analyzed using the Lagrangian formalism, defined as
\begin{equation}
\mathcal{L}=\frac{1}{2}g_{\mu\nu}\frac{\mathrm{d} x^{\mu} }{\mathrm{d} \sigma}\frac{\mathrm{d} x^{\nu} }{\mathrm{d} \sigma}.
\label{largangiian}
\end{equation}
Here, \( x^\mu = (t, r, \theta, \phi) \) denote the spacetime coordinates, and \( \sigma \) is an affine parameter along the geodesics. The conjugate momenta are defined as
\begin{equation}
p_\mu = \frac{\partial \mathcal{L}}{\partial \dot{x}^\mu}.
\end{equation}
The dot denotes differentiation with respect to the affine parameter \( \sigma \). Without loss of generality, we restrict the motion to the equatorial plane by setting \( \theta = \frac{\pi}{2} \) and \( \dot{\theta} = 0 \). 

The spacetime symmetries lead to two Killing vectors, \( \partial_t \) and \( \partial_\phi \), which imply the existence of two conserved quantities:
\begin{equation}
p_t = -E, \quad p_\phi = L.
\end{equation}
Here, \( E \) and \( L \) represent the conserved energy and angular momentum of the photon, respectively \cite{shapiro2008black, Chabab:2019kfs}. The equations of motion associated with the cyclic coordinates \( t \) and \( \phi \), obtained from the Euler--Lagrange equations, are given by
\begin{align}
\frac{\mathrm{d} t}{\mathrm{d} \sigma}&=\frac{E}{f(r)},
\label{motioneuqpt}\\
\frac{\mathrm{d} \phi}{\mathrm{d} \sigma}&=\frac{L}{r^{2}}.
\label{motioneuqpphi}
\end{align}

We now introduce the effective potential by using the normalization condition for null geodesics, \( g_{\mu\nu} \dot{x}^\mu \dot{x}^\nu = 0 \). Taking this constraint into account, the radial equation of motion can be written as
\begin{equation}\label{Hamilton}
\dot{r}^{2}+V_{\text{eff}}(r)=0.
\end{equation}
The effective potential \( V_{\text{eff}}(r) \) is given by
\begin{equation}
\frac{V_{\text{eff}}(r)}{E^2}= \xi^2 \frac{f(r)}{r^2} - 1,
\label{veff}
\end{equation}
where \( \xi = \frac{L}{E} \) denotes the impact parameter. According to \cite{Chandrasekhar:2018sjg}, photon motion is allowed only in regions where \( V_{\text{eff}}(r) \leq 0 \). We therefore focus on unstable circular photon orbits.
In such orbits, a photon arriving from infinity reaches a turning point with vanishing radial velocity. The radius \( r_{ps} \) of the unstable circular orbit (photon sphere) is determined by the conditions
\begin{equation}
V_{\text{eff}}(r_{ps}) = 0, \quad 
\left.\frac{\partial V_{\text{eff}}}{\partial r}\right|_{r=r_{ps}} = 0, \quad 
\left.\frac{\partial^2 V_{\text{eff}}}{\partial r^2}\right|_{r=r_{ps}} < 0.
\label{condveff}
\end{equation}
Using the first condition in Eq.~\eqref{condveff} together with Eq.~\eqref{veff}, the critical impact parameter corresponding to unstable circular orbits is given by
\begin{equation}
\xi_0 = \frac{r_{ps}}{\sqrt{f(r_{ps})}}.
\label{Xi0}
\end{equation}

\begin{figure}[h]
	\centering 
	\includegraphics[scale=0.5]{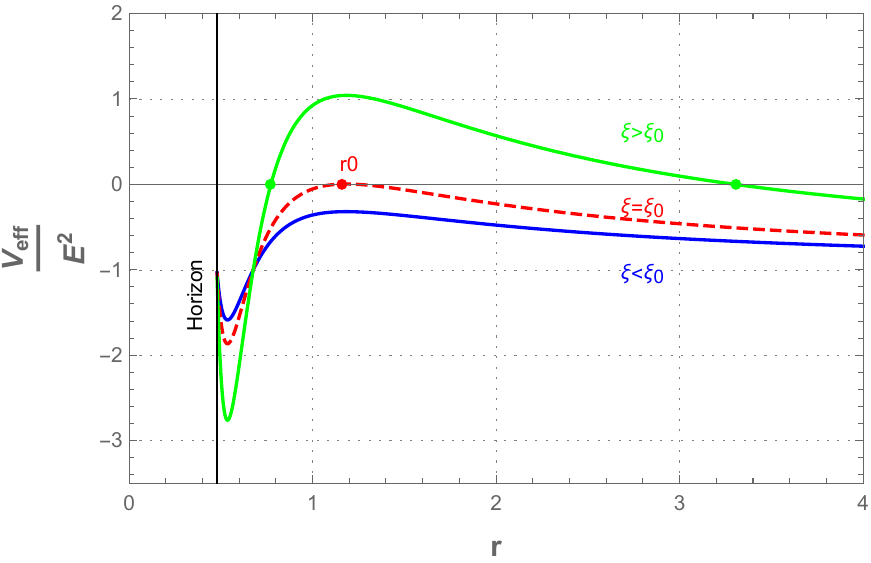}  
	\caption{
		Effective potential for photon motion around a charged black hole in Rastall--massive gravity, showing the existence of unstable circular photon orbits. The red and green points indicate turning points. Parameters are set to \( S=0.72 \), \( \alpha=0.02 \), \( \beta=1.6 \), \( a=5.29 \), \( b=1 \), and \( Q=0.5 \).
}
	\label{vefffig}
\end{figure} 
Figure~\ref{vefffig} illustrates the possible trajectories of photon motion. The turning points are determined by the condition \( V_{\text{eff}}=0 \). Depending on the value of the impact parameter \( \xi \), three distinct regimes can be identified:
\begin{itemize}
	\item For \( \xi > \xi_0 \): photons coming from infinity reach a turning point and are scattered along hyperbolic trajectories.
	\item For \( \xi = \xi_0 \): this corresponds to the critical case, separating capture and scattering. Photons asymptotically approach the unstable circular orbit.
	\item For \( \xi < \xi_0 \): photons are captured by the black hole.
\end{itemize}
The corresponding photon capture cross-section is given by
\begin{equation}
\sigma_{\text{capture}} = \pi \xi_0^2 = \frac{\pi r_{ps}^2}{f(r_{ps})}.
\label{grosssection}
\end{equation}

\subsection{Thermodynamic phase transitions}
	
In this section, we establish a connection between thermodynamic phase transitions and unstable circular photon orbits. In particular, we extend and generalize the analysis developed in \cite{Chabab:2019kfs, Wei:2017mwc} to the case of charged black holes in Rastall--massive gravity. To this end, we substitute the metric function given in Eq.~\eqref{metric} into the effective potential in Eq.~\eqref{veff}, and apply the second condition in Eq.~\eqref{condveff}. This leads to an algebraic equation for the photon sphere radius \( r_{ps} \), which depends on the Rastall parameter \( \beta \). For fixed values of \( \beta \), the constraint equation takes the form
\begin{equation}
4 Q^2 + r_{ps} \left[-6M + r_{ps} \left(2 + 2 b + r_{ps} (a + \alpha)\right) \right]=0.
\label{photon_constraint}
\end{equation}
Solving this equation allows one to determine the photon sphere radius \( r_{ps} \), which can then be substituted into Eq.~\eqref{Xi0} to obtain the corresponding critical impact parameter \( \xi_0 \).

\begin{figure}[h]
	\centering 
	\includegraphics[scale=0.45]{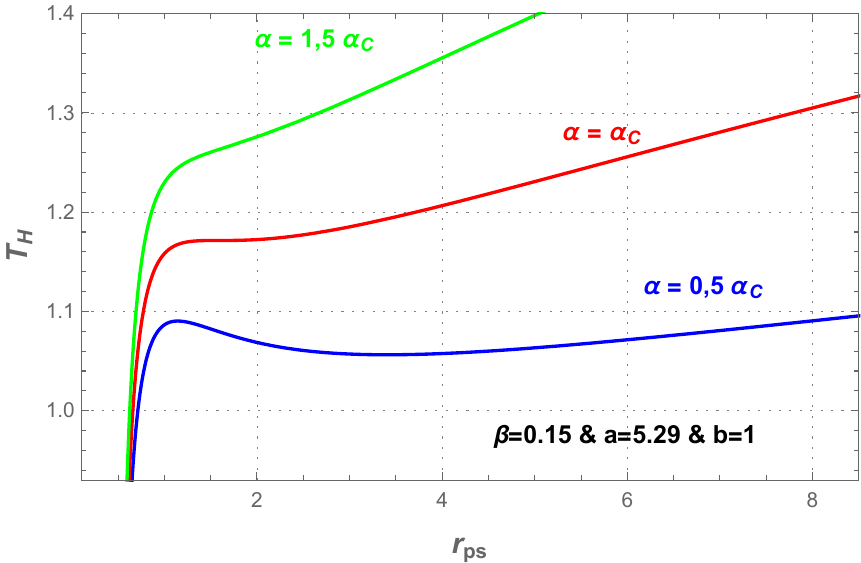}   
	\includegraphics[scale=0.45]{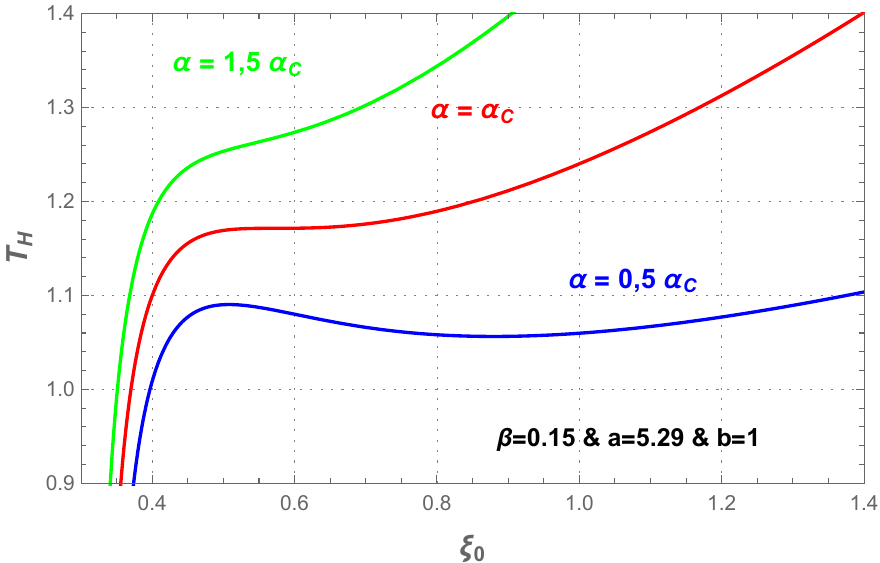} \\ 
	\includegraphics[scale=0.45]{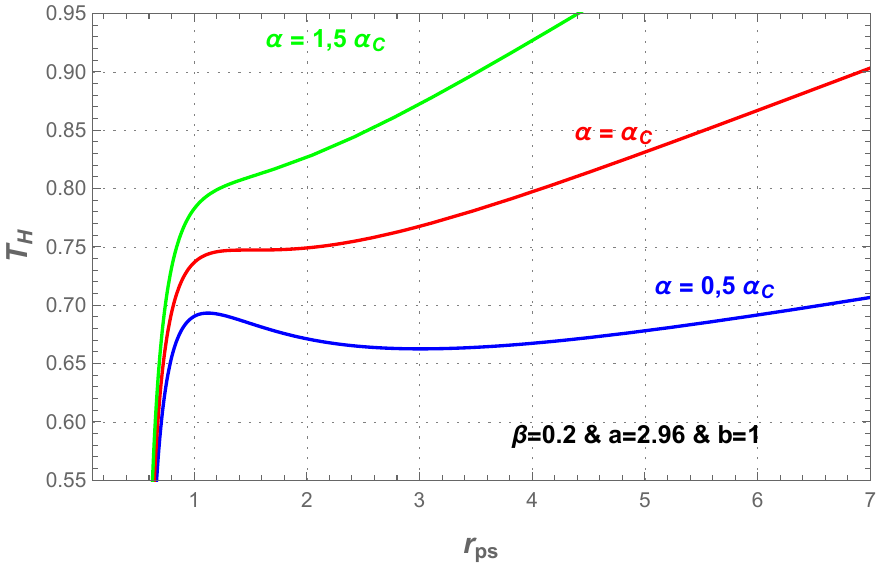}   
	\includegraphics[scale=0.45]{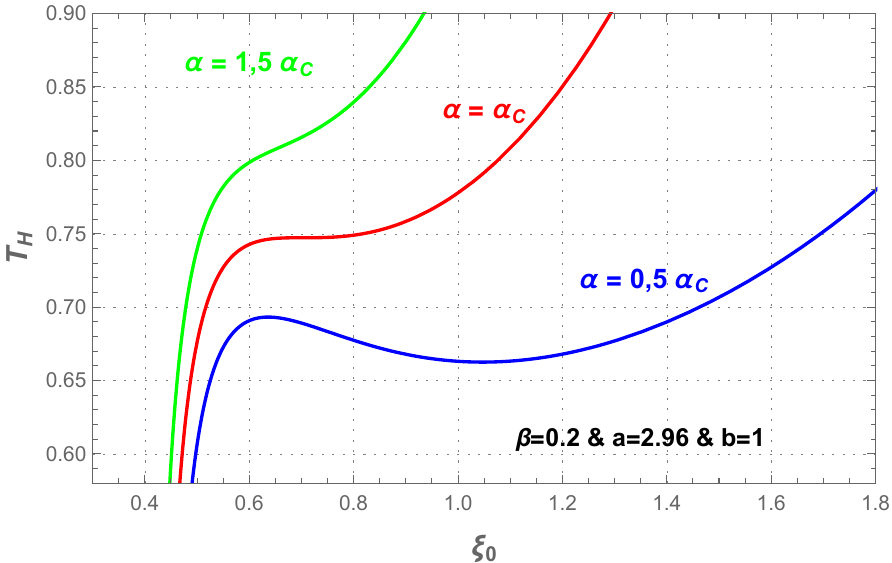}  \\
	\includegraphics[scale=0.45]{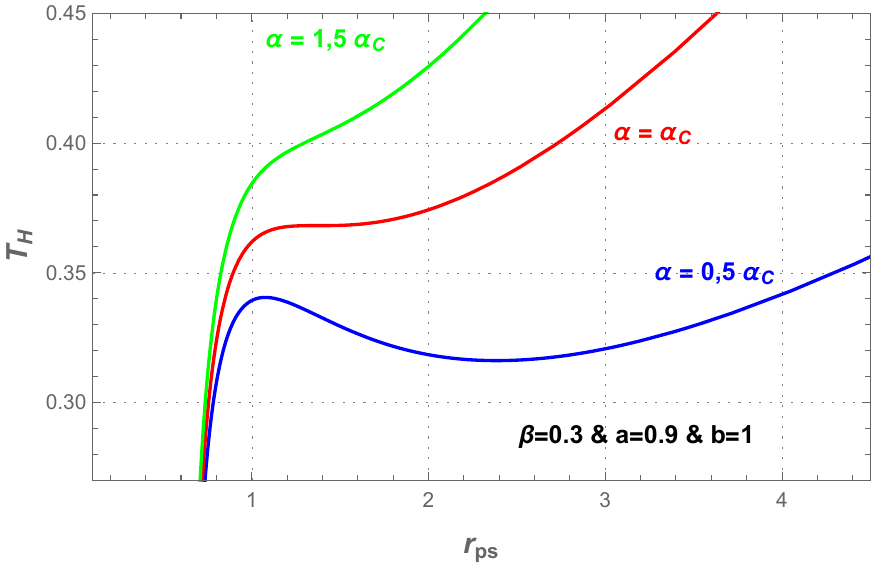}   
	\includegraphics[scale=0.45]{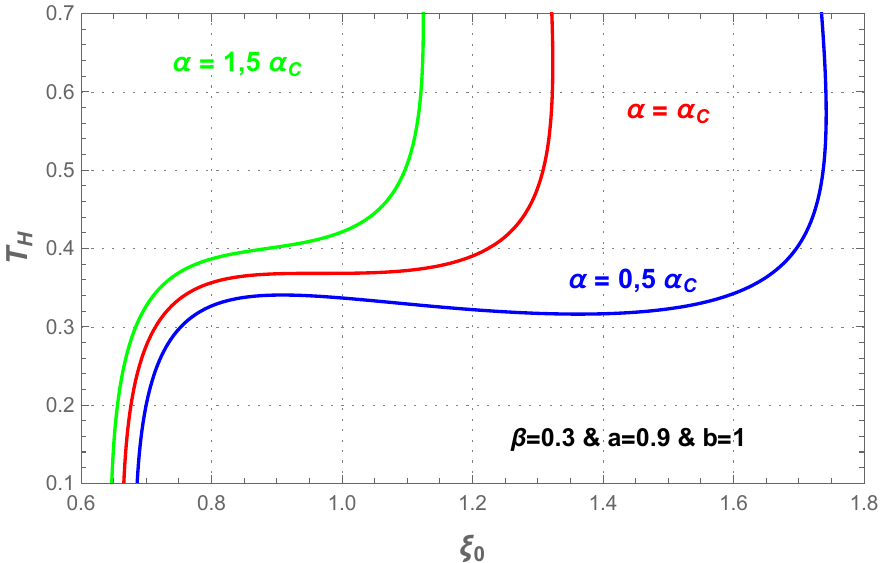}
	\caption{\(T_H - r_{ps}\) (left panels) and \(T_H - \xi_0\) (right panels) diagrams for different values of the Rastall parameter \( \beta \) and the massive gravity parameter \( a \), illustrating the thermodynamic phase transition through photon sphere observables.}
	\label{Fig6}
\end{figure}

In the following, we present the results shown in Fig.~\ref{Fig6}, where the thermodynamic is analyzed in terms of the photon sphere.
The figure shows the behavior of the Hawking temperature as a function of the photon sphere radius \( r_{ps} \) and the impact parameter \( \xi_0 \). For \( \alpha < \alpha_c \), both sets of curves exhibit a non-monotonic behavior with an oscillatory structure, which is characteristic of a first-order phase transition between small and large black holes.

At the critical point \( \alpha = \alpha_c \), the curves develop an inflection point, signaling a second-order phase transition. For \( \alpha > \alpha_c \), the curves become monotonic, indicating the absence of phase transitions. Remarkably, this behavior is entirely analogous to that observed in the \( (T_H - r_h) \) and \( (T_H - S) \) diagrams. Therefore, the phase structure revealed through the photon sphere radius and the impact parameter is consistent with that obtained from the thermodynamic variables. These results demonstrate that the photon sphere radius \( r_{ps} \) and the critical impact parameter \( \xi_0 \) encode equivalent signatures of phase transitions as the horizon radius and entropy. Consequently, unstable circular photon orbits provide an alternative and powerful probe of small/large black hole phase transitions in Rastall--massive gravity.
	
	\section{Geometrothermodynamics}
	
	We investigate the thermodynamic geometry within the framework of Rastall--massive gravity in the presence of a quintessence field. Our aim is to show that the critical behavior is encoded both in the divergences of the heat capacity and in the singularities of the curvature scalar associated with the Quevedo metric,
	
	Quevedo \cite{Quevedo:2006xk,Quevedo:2008xn} introduced a thermodynamic metric constructed from a Legendre-invariant structure defined on a \(\mathcal{T}^{2n+1}\)-dimensional phase space, based on a thermodynamic potential and its derivatives. In this formalism, one introduces a coordinate system \(\{\Phi, E^i, I^i\}\), where \( \Phi \) is the thermodynamic potential, while \( E^i \) and \( I^i \) denote the extensive and intensive variables, respectively.
	
	In the equilibrium space, the Quevedo metric is defined as
	\begin{equation}
	g^{Quv} = \left(E^k \frac{\partial \Phi}{\partial E^k}\right)\left(\eta_{ij} \, \delta^{jk} \, \frac{\partial^2 \Phi}{\partial E^k \partial E^l} \, dE^i dE^l \right),
	\end{equation}
	where \( \frac{\partial \Phi}{\partial E^k} = \delta_{kj} I^j \).
	
	By considering the graviton mass \( m_g \) as an extensive thermodynamic variable, one can redefine the massive gravity parameters \( a \) and \( b \) as
	\[
	a = \gamma m_g^2, \qquad b = \xi m_g^2,
	\]
	where \( \gamma \) and \( \xi \) are constant parameters. Accordingly, the black hole mass formula, given in Eq.~\eqref{mass}, can be rewritten as:

		\begin{equation}
	M(s,m_g) = \frac{\sqrt{\pi}\left( Q^2 + \alpha\, \pi^{\frac{3}{2(1-\beta)}} s^{\frac{3}{2(1-\beta)}}  + \frac{s \left[ 1 + m_g^2 \left( \frac{\sqrt{s} \gamma}{\sqrt{\pi}} + \xi \right) \right]}{\pi} \right)}{2\sqrt{s}}
	\label{massmg}
	\end{equation}
	Here, \( S = \pi r_h^2 \) denotes the black hole entropy.

In this context, to simplify the calculations within the scope of this work, we consider a 7- dimensional phase space whose coordinates are \(Z_A = \{M, S, T, m_g, Q, \phi, \mu\}\), where M is chosen as a thermodynamic potential, while S, \(m_g\), and Q are the extensive variables associated with the intensive variables T, \(\mu\) and \(\phi\) respectively. Consequently, the fundamental Gibbs formula is expressed as follows:
\begin{equation}
    \Phi = dM - TdS - \phi dQ - \mu dm_g
    \label{Gbbisform}
\end{equation}
where $M$ is selected as the thermodynamic potential.

According to \cite{Hendi:2015xya,Hendi:2015rja,Chabab:2019mlu,Soroushfar:2019ihn,EslamPanah:2018ums}, the Quevedo metric can be expressed in the following form:
	\begin{equation}
	ds^2 = (ST + \phi Q +\mu m_g)(-M_{SS}dS^2 + M_{QQ}dQ^2 + M_{m_g m_g}dm_g^2  ).
	\label{quevedometr}
	\end{equation}
Here, \( \phi=\left.\frac{\partial M}{\partial Q}\right|_{S,m_g,\alpha,\beta} \) is the electric potential, while \( \mu=\left.\frac{\partial M}{\partial m_g}\right|_{S,Q,\alpha,\beta} \) is the thermodynamic quantity conjugate to the graviton mass.	The components of the Quevedo metric are readily obtained from Eq.~\eqref{massmg}, combined with Eq.~\eqref{quevedometr} and the entropy expression,
\begin{align}
    g_{SS} = - \frac{C \times D}{32 \pi^{3/2} s^3 (\beta - 1)^3},\qquad
    g_{QQ} = \frac{D}{4 \sqrt{\pi} s (\beta - 1)},\qquad
    g_{m_g m_g} = \frac{\left( \sqrt{s} \gamma + \sqrt{\pi} \xi \right) D}{4 \pi^2 (\beta - 1)}
\end{align}
where the terms \(C\), \(D\) and \(\delta\) are given by:
   \begin{align}
   C &= 3 \alpha \beta \pi^{1 + \delta} s^{-\delta} (\beta + 2) - (\beta - 1)^2 \left( -3 \pi Q^2 + s + m_g^2 s \xi \right),\qquad \delta = \frac{3}{2(\beta - 1)},\nonumber\\
    D &= -\alpha \pi^{\delta\beta} s^{-\delta} (\beta + 2) + (\beta - 1) \left( 3 \pi^{3/2} Q^2 + 6 m_g^2 s^{3/2} \gamma + \sqrt{\pi} (s + 5 m_g^2 s \xi) \right).
  \end{align}
While remaining components of the Quevedo metric vanish.
	
We have calculated the curvature scalar associated with the Quevedo metric. Though the expression is lengthy, it can be written as a general ratio of two functions,
\begin{equation}
R^{Qu}=\frac{h(S,Q,\beta,\alpha,m_g)}{g(S,Q,\beta,\alpha,m_g)}.
\end{equation}

where $h$ and $g$ denote the numerator and denominator of the Quevedo metric respectively.  

\begin{figure}[h]
	\centering 
	\includegraphics[scale=0.45]{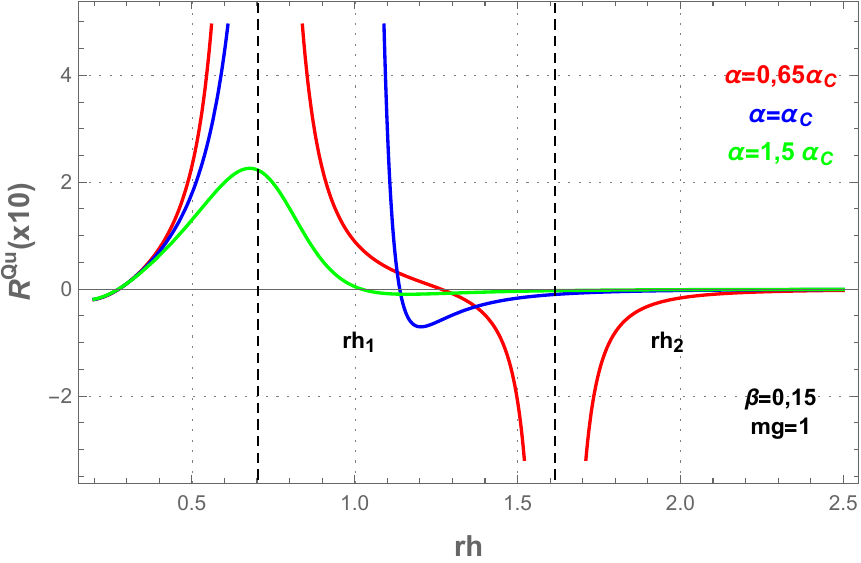}   
	\includegraphics[scale=0.45]{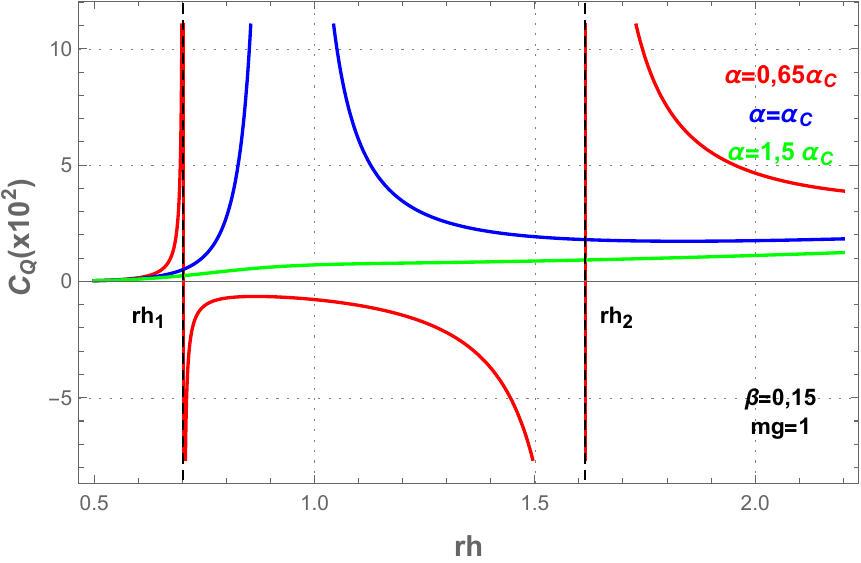} \\ 
	\includegraphics[scale=0.45]{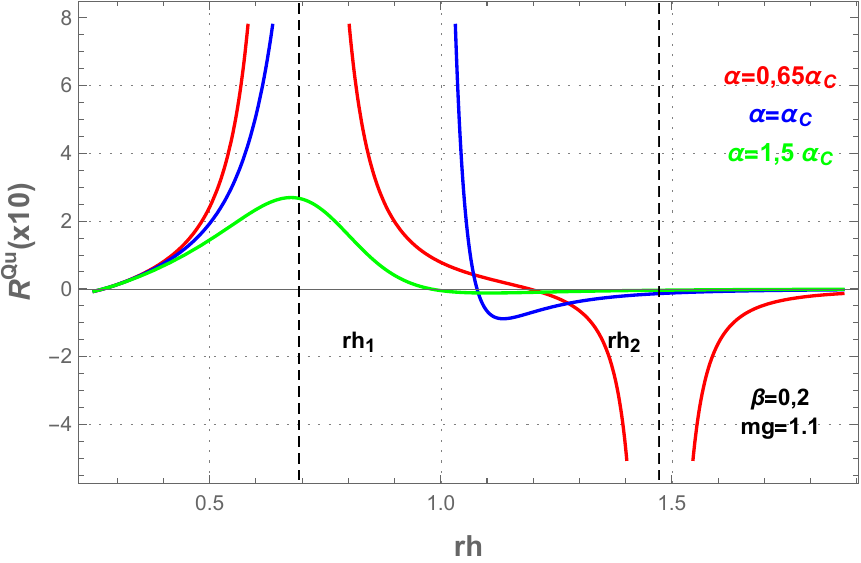}   
	\includegraphics[scale=0.45]{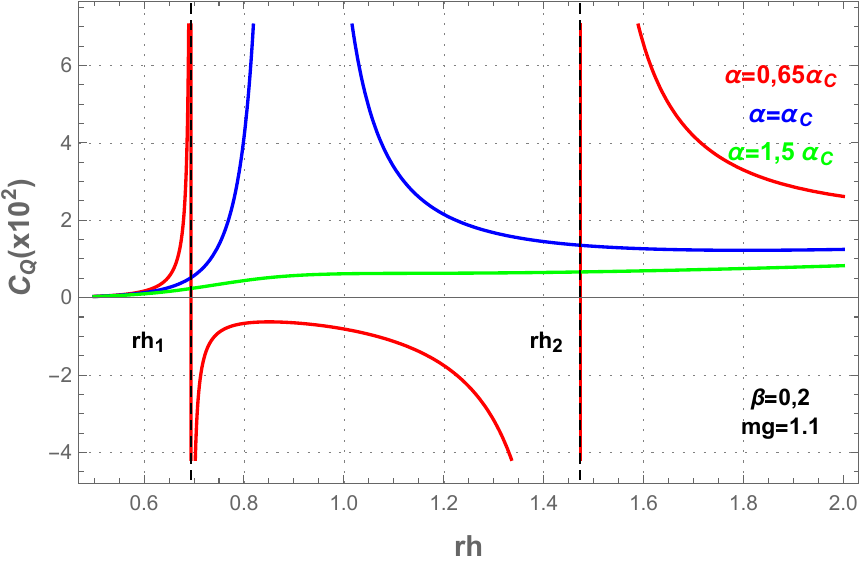}  \\
	\includegraphics[scale=0.45]{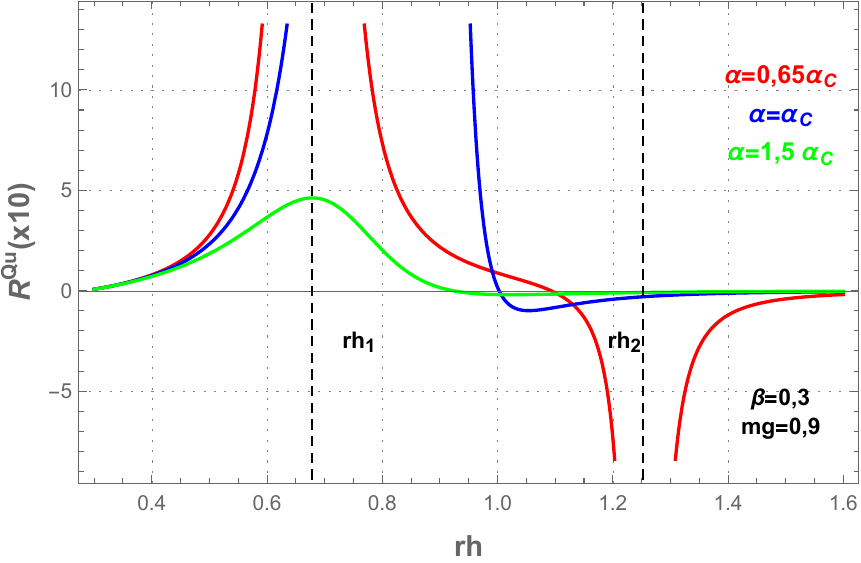}   
	\includegraphics[scale=0.45]{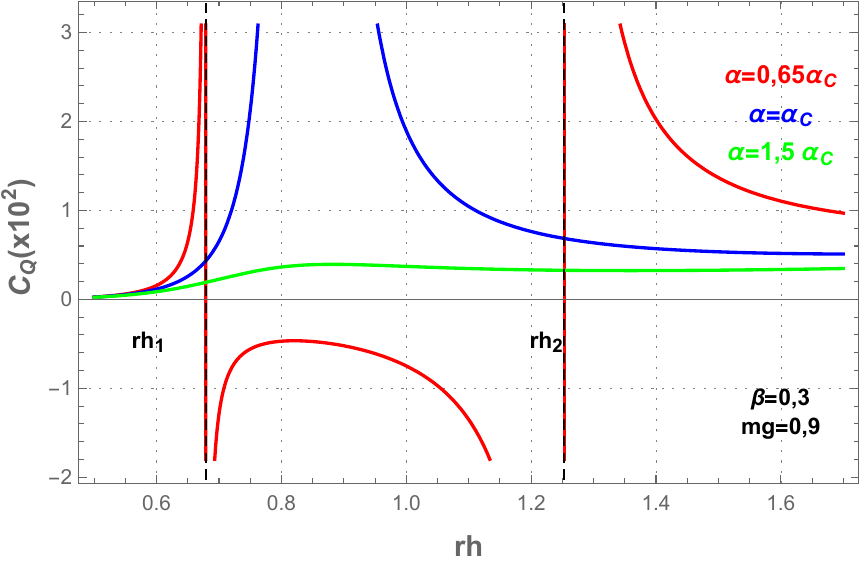}
	\caption{\(R^{Qu} - r_h\) diagrams (left panels) and \(C_Q - r_h\) diagrams (right panels) for different values of the parameters. We set \( \gamma=1 \) and \( \xi = 1.24,\, 1,\, 0.827 \) from top to bottom.}
	\label{geother}
\end{figure}

Figure~\ref{geother} displays the behavior of the heat capacity and the Quevedo curvature scalar. The values of the graviton mass \( m_g \) and the parameter \( \xi \) are chosen such that \( b=1 \), and the same critical values as those listed in Table~\ref{table1} are reproduced. This allows us to fix the universal ratio \( \frac{\alpha_c \nu_c}{T_c} = \frac{3}{8} \), and to compare our results with those previously established.
The vertical lines \( r_{h1} \) and \( r_{h2} \) indicate the divergence points of the heat capacity. In Fig.~\ref{geother}, we plot both the heat capacity and the Quevedo curvature scalar as functions of the horizon radius.

For \( \alpha < \alpha_c \), the Quevedo curvature scalar exhibits two divergence points, which exactly coincide with those of the heat capacity at \( r_{h1} \) and \( r_{h2} \). This behavior signals a phase transition between small and large black holes. The change of sign of the curvature scalar near these divergency points reflects a change in thermodynamic stability.
For \( \alpha = \alpha_c \), the Quevedo curvature scalar shows a single divergency, corresponding to the critical point.
For \( \alpha > \alpha_c \), the curvature scalar exhibits a smooth behavior with a local maximum but no divergency, indicating the absence of phase transitions and the stability of the black hole configuration.

Overall, these results demonstrate that thermodynamic geometry reproduces the same phase structure as standard thermodynamics. In particular, the singularities of the Quevedo curvature scalar encode the same information as the divergences of the heat capacity for charged black holes in Rastall--massive gravity in the presence of a quintessence field. This confirms the consistency between thermodynamic geometry approach and standard thermodynamic analysis.

While the thermodynamic metric is formally constructed in the \({S,Q,\beta,\alpha,m_g}\) phase space, the Quevedo metric \(R^{Qu}\) inherits the full parametric dependence of the fundamental mass \(M(S,Q,\beta,\alpha,m_g)\). Consequently, the critical behavior associated with the quintessence parameter \(\alpha\) is directly mapped onto the singularities of \(R^{Qu}\). This ensures a rigorous correspondence between thermodynamic phase transitions and the underlying geometry of the equilibrium monifold.

	\section{Conclusion}
	\label{Conclusion}
	
	In this paper, we have derived and analyzed a new charged black hole solution in the framework of the combined massive and Rastall gravity, surrounded by a quintessence field in asymptotically flat spacetime.  Several limiting cases have been examined with the aim to scrutinize the specific role of the model parameters. \\
	
	The integration constant associated with the structural properties of the quintessence field is interpreted as a thermodynamic variable. We show that, as the Rastall parameter \( \beta \) increases, both the massive gravity parameter \( a \) and the critical value \( \alpha_c \) decrease. For positive values of \( \beta \), both \( a \) and \( \alpha_c \) remain positive, analysing, ensuring that the universal ratio \( \frac{\alpha_c \nu_c}{T_c} = \frac{3}{8} \) is only satisfied.\\
	
	In contrast, for negative values of \( \beta \), consistency requires negative values of the parameters, which leads to the disappearance of physically meaningful phase transitions. Overall, the thermodynamic behavior closely resembles that of the charged RN--AdS black hole with thermal features analogous to those of a Van der Waals fluid.\\
	
	In the second part of this work, we have investigated phase transitions from the perspective of unstable circular photon orbits within the Rastall--massive gravity framework. Starting from the null geodesic equations, we determined the photon sphere radius and the corresponding critical impact parameter. By analyzing the \( T\text{--}r_{ps} \) and \( T\text{--}\xi_0 \) diagrams, we found that they reproduce exactly the same phase transition structure observed in the \( T\text{--}r_h \) and \( T\text{--}S \) diagrams.\\
	
	To summarize, thanks to the effective contribution of the quintessence parameter, the phase structure of charged black holes in Rastall--massive gravity is fully consistent with that of RN--AdS black holes, even when an explicit cosmological constant is omitted. This supports the interpretation of the parameter \( \alpha \) as a thermodynamic state variable analogous to pressure. These results suggest that photon sphere observables, potentially accessible through gravitational lensing or black hole shadow measurements, could provide a noteworthy indirect observational signature of  black hole thermodynamic phase transitions.\\

    Finally, we have conducted a further analysis employing the thermodynamic geometry with the Quevedo metric in the  Rastall--massive gravity framework in presence of  quitessential field.  We have first defined the extensive parameters and computed the various components. The obtained results showed a critical behaviour perfectly similar to that of standard thermodynamics, where the phase structure and divergent points are fully consistent with those produced through the heat capacity and Gibbs free energy.

\end{document}